\documentclass[11pt]{article}
\usepackage{blois,epsfig}

\def\be{\begin{equation}}
\def\ee{\end{equation}}
\def\bea{\begin{eqnarray}}
\def\eea{\end{eqnarray}}

\begin{document}
\vspace*{4cm}
\title{Orientation of the galaxy groups in the Local Supercluster}
 
\author{Piotr Flin}
\address{Jan Kochanowski University, Institute of Physics, ul. Swietokrzyska 15,
25-406 Kielce, Poland}
 
\author{W{\l}odzimierz God{\l}owski}
\address{Uniwersytet Opolski, Institute of Physics, ul.  Oleska  48,
45-052 Opole, Poland}

\maketitle\abstracts{
The paper discusses the problem of the orientation of galaxies in groups
in the Local Supercluster (LSC). The existence of the
preferred orientation of galaxy group is shown. We found that the orientation 
 of  galaxy groups in the
Local Supercluster in the scale till about 20 Mpc is  strongly  correlated
with the distribution of  neighbouring groups. The line joining the two
brightest galaxies is in alignment  with  both the group major axes and
the direction toward  the  centre  of  the  LSC, i.e. Virgo  cluster. These
correlations suggest that two brightest galaxies were formed in filaments
of matter directed towards  the  protosupercluster centre. Afterwards, the
hierarchical clustering leads to aggregation of galaxies around these two
galaxies. The groups are formed on the same or similarly oriented filaments.
This picture is in agreement with the predictions of numerical simulations.}

\section{Introduction}
 
Starting from Binggeli paper \cite{a1}, several authors studied the orientation
of galaxy groups and clusters \cite{a2,a3,a4,a5,a6,a7,a8,a9,a10}. On
the basis of optical and X-ray data they found that structures exhibit
a tendency to be orientated toward their neighbours. The interpretation
of this effect has changed in the last thirty years, but the main idea
that this should reflect  conditions during the structure formation is
still very popular. Numerical simulations
\cite{a11,a12,a13,a14,a15,a16,a17,a18,a19,a20,a21} gave a better
understanding of physical processes leading to structure formation.
These simulations were performed in the framework of the cold dark
matter(CDM) model, presently regarded as the correct description of
the large scale structure formation.

Several  numerical simulations, using different approaches  and
codes, led to the conclusion that the preferred orientation  of
galaxy  clusters in the CDM model is a natural  consequence  of
processes  leading to structure formation due to  gravitational
interaction along filamentary structures. In order to  confirm,
or  deny,  this  scheme of structure origin  we  carry  out  an
analysis   of  the  Local  Supercluster  (LSC)  galaxy   groups
alignment.  Groups  were  taken from the  Catalogue  of  Nearby
Galaxies \cite{a22}. We selected structures having at least 10
members. There are 61 such groups.
 
\begin{figure}
\hskip 0.5cm
\psfig{figure=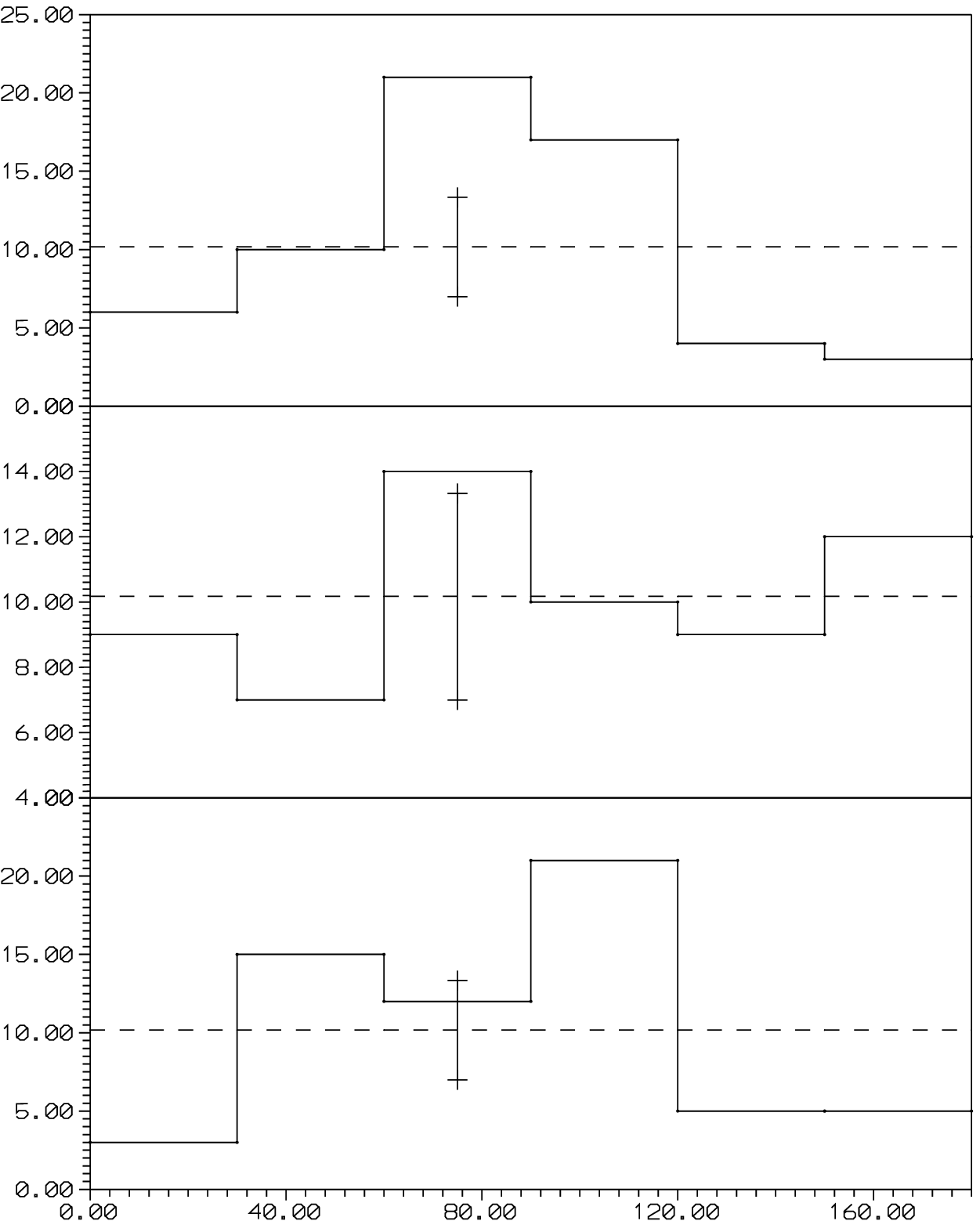,height=3.6in}
\psfig{figure=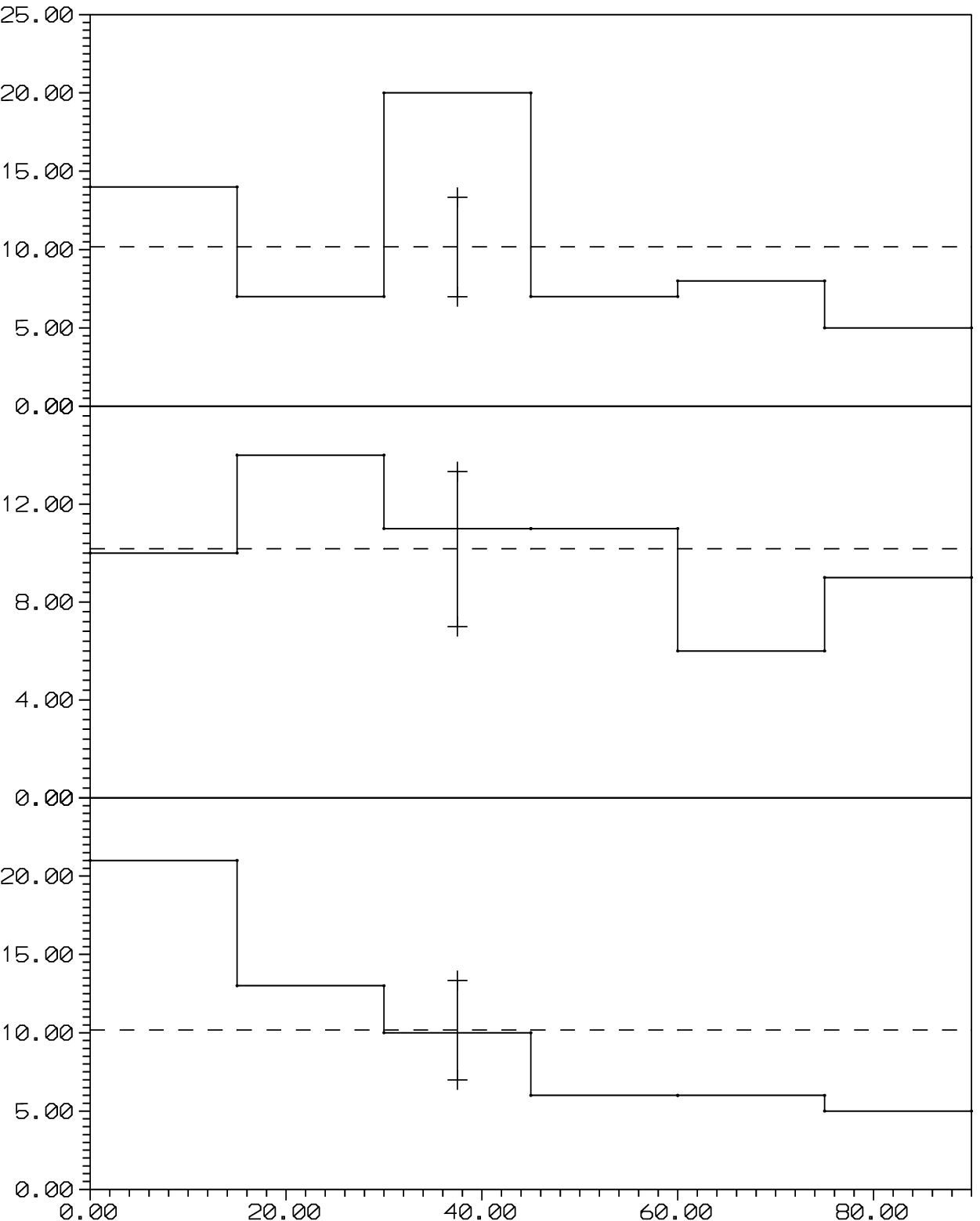,height=3.6in}
\caption{The distribution (from top to bottom) of the position
angles $PA_g$, $PA_l$  $PA_V$ (left panel) and the distribution 
(from top to bottom) of the differences between position angles
$PA_g-PA_V$, $PA_l-PA_V$, $PA_g-PA_l$ (right panel).
\label{fig1}}
\end{figure}

\section{Observational data and analysis}
 
It  was assumed that groups are two axial ellipsoids. The shape
of   each  group  has  been  determined  considering  only  the
projected position of galaxies on the celestial sphere  in  the
supergalactic   coordinate  system  L,  B  and   applying   the
covariance  ellipse method. This procedure gives  the  position
angle of the group major axis.
 
The   position   angle  of  each  group  $PA_g$  is   calculated
counterclockwise  from  the great circle  passing  through  the
position of the cluster centre on the celestial sphere and  the
northern pole of the LSC. It was assumed that the location of a
group centre corresponds to the mean of
L  and  B coordinates of member galaxies and the mean of radial
velocity,  as  given in the Catalogue. Using standard  formulae
from  spherical  trigonometry,  we  calculated  the  directions
between  the  centre  of  each group and  the  centres  of  the
remaining groups. Each direction is a part of the great  circle
joining the centres of two groups. For each group we calculated
the  acute angle $\phi$ between the position angle of the major axis
of  a given group $PA_g$ and direction towards other groups.  We  also
investigated the alignment of the brightest group galaxies  and
the  parent  group.  We  determined the position  of  the  line
joining  two brightest galaxies in the group $PA_l$ and  checked
the orientation of this line relative to the position angle  of
the  parent  group $PA_g$, the position angle of  the  brightest
galaxy $PA_{bm}$ and the direction towards Virgo cluster $PA_V$.
 
\begin{figure}
\hskip 3.0cm
\psfig{figure=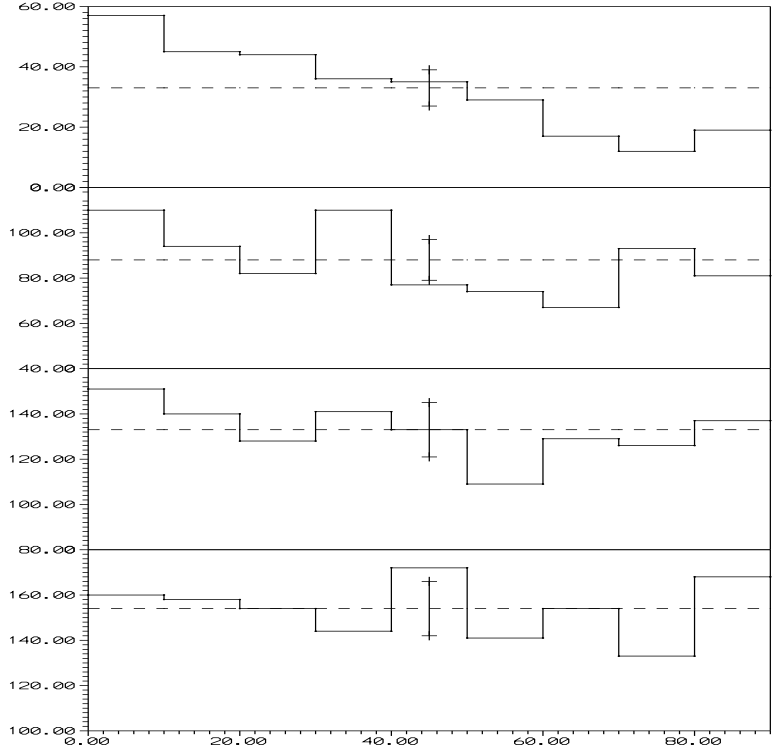,height=3.9in,width=4.0in}
\caption{The distribution of the acute angle $\phi$  between the
position angle of the major axis of a given group ($PA_g$) and
direction towards other groups. From top to bottom the
distributions for groups with
$D \leq 10 Mpc$, $10<D\leq 20 Mpc$, $10<D \leq 20 Mpc$
and $D>20 Mpc$ are presented respectively.
The dashed  lines denoted the isotropic distribution. 
\label{fig2}}
\end{figure}

We checked for isotropy all discussed distributions of position
angles $PA_g$, $PA_l$, $PA_{bm}$, $PA_V$ having range $0^o-180^o$,
as well as differences between  position angles $PA_g-PA_V$, $PA_l-PA_V$,
$PA_g - PA_l$, $PA_{bm} - PA_g$, $PA_{bm} - PA_l$, $PA_{bm} - PA_V$
being the acute angles (fig.1). It was done using the Kolmogorov - Smirnov
test and the $\chi^2$ test. Additionally, we carried out the analysis
of these angles for 35 galaxy groups having at least 20 members.

\section{Results}
 
The  distribution of position angles of the brightest  galaxies
($PA_{bm}$)  is  isotropic as is observed in galaxy  structures  not
containing   cD  galaxy,  which  is  the  case  of   LSC.   The
distributions  of  group position angles  $PA_g$  and  the  line
joining  two  brightest  galaxies  $PA_l$  are  anisotropic   at
confidence level 95\% In the case of the $\chi^2$ test the $PA_l$
is anisotropic only when 35  richer groups  are  analysed.  The
distribution   of  direction  toward  Virgo   centre   $PA_V$ is
anisotropic at the confidence level 95\%.  The strong excess
of  position  angles $PA_g$ is observed in the bin $80^o-100^o$,
which corresponds to the location of the supergalactic equator.
In this bin, the excess of position angles of the structures is
$5\sigma$, while the excess of the position line joining two brightest
galaxies $PA_l$  is  $2.5\sigma$ (for  35  galaxy  groups  only),  when
compared to the number expected in a random distribution.
 
The $\chi^2$ test shows  that  the difference  between  the  group
position angle $PA_g$ and direction towards Virgo cluster  $PA_V$
is  not random at the confidence level 99\% or 95\% in  the
case  of richer and poorer groups respectively. For 41 clusters
the  differences $PA_g-PA_V$ are less than $45^o$, while only for 20
clusters  they  are greater than $45^o$. The distribution of the
difference  $PA_V-PA_l$ is anisotropic only for richer groups. The
difference  of  angles $PA_g-PA_l$ is strongly anisotropic  at  the
confidence level 99\%.  The observed excess is below $45^o$.
 
The  structures have the tendency to point each other  only  in
the  case when the distance between groups is smaller  than  $20 Mps$
($H_0=75 km\, s^{-1} Mps^{-1}, q_0 =1/2$ fig.2). We obtain the anisotropy
for the differences between the osition angle $PA_g$ and direction
towards other groups.  This effect is at the level of $7 \sigma$ for 
$D\leq 10 Mpc$ and at $2.5\sigma$ for $10<D \leq 20 Mpc$. For the  
sample of 35 richer galaxy groups, we also obtained similar effect, 
but at $2.5\sigma$ level and only for $D \leq 10 Mpc$. A similar 
tendency, but at $2.3 \sigma$ level is observed for the difference 
between the line joining two brightest galaxies ($PA_l$) and direction 
towards other groups. Moreover, anisotropies  are  noted  only in the
coordinate system connected with the LSC and they are absent in the 
equatorial coordinate system, which gives further evidences that the 
orientation is connected with the LSC itself.
 
\section{Conclusions}
 
From the presented analysis of the orientation of galaxy groups
in   the  Local  Supercluster  the  following  picture  of  the
structure  formation appears. The two brightest  galaxies  were
formed  first.  They  originated in the  filamentary  structure
directed  towards the centre of the protocluster. This  is  the
place where the Virgo cluster centre is located now.

Due  to gravitational clustering, the groups are formed in such
a  manner that galaxies follow the line determined by  the  two
brightest   objects.  Therefore,  the  alignment  of  structure
position  angle  and  line joining two  brightest  galaxies  is
observed.  The other groups are forming on the same  or  nearby
filament.  The flatness of the LSC additionally contributes  to
the  observed alignment of galaxy groups. The majority  of  the
groups  lie  close  to  us. From the selection  effect  in  the
catalogue  the  lack of groups further than the  Virgo  Cluster
centres is observed.

This  picture is in agreement with predictions of  several  CDM
models,  in  which structure formation is due  to  hierarchical
clustering.   Moreover,  the  formation  is  occurring  on  the
filamentary structure.

\section*{Acknowledgments}
 This work was partially suported by the Jan Kochanowski University
 grant BS 052/09.

\end{document}